\documentstyle[aps,preprint]{revtex}

\begin{document}
\draft

\title
{\bf Ginzburg-Landau Expansion and Physical Properties of
Odd-Gap Superconductors}
\author{E.Z.Kuchinskii,\ A.I.Posazhennikova and M.V.Sadovskii}
\address
{Institute for Electrophysics,\ Russian Academy of Sciences,\ Ural Branch,\ \\
Ekaterinburg,\ 620219, Russia\\
E-mail:\ sadovski@ief.e-burg.su}
\maketitle

\begin{center}
{\sl Submitted to JETP}
\end{center}

\begin{abstract}
Ginzburg-Landau expansion is derived for superconductors with the gap-function
odd over $k-k_{F}$. It is shown that in this odd-gap case Ginzburg-Landau
coefficients possess an additional dependence on the pairing coupling constant
which leads to the appropriate dependence of the physical properties.
Non-magnetic impurities influence the odd-gap pairing in a significant way
and the mechanism of this influence is quite different from that predicted by
the usual theory of "dirty" superconductors. In particular this is reflected
in an anomalous behavior of the upper critical field $H_{c2}$ slope close to
$T_{c}$. We also derive the complete temperature dependence of $H_{c2}$ for
the odd-gap case.
\end{abstract}

\newpage
\narrowtext

\section{Introduction}
Recently Mila and Abrahams\cite{MA} have proposed an interesting model of
superconducting pairing with the gap function odd over the parameter
$k-k_{F}$ (i.e.\ quasiparticle energy $\xi_{k}=v_{F}(|\vec k|-k_{F})$,
with respect to the Fermi level). In this model the BCS pairing state is
possible even for the arbitrarily strong repulsion between the electrons.
This odd-gap state is realized for the strong enough repulsion when the usual
``even'' superconducting state is destroyed while the attractive part of the
pairing interaction is also strong enough (the pairing coupling constant is
greater than some critical value\cite{MA}).
Naturally, \ this model seems attractive as a possible explanation of the
superconducting state in strongly-correlated systems,\ though it is clear now
what it is an unlikely candidate for high-temperature superconducting oxides,\
e.g.\ because of the assumed isotropic nature of pairing (it is well
established now experimentally that the pairing state in oxides is either
$d$-wave or anisotropic $s$-wave).
At the same time, this odd-gap model\cite{MA} is interesting itself as a model
of a possible new ``exotic'' superconducting state with the properties quite
different from traditional superconductors.
Thus it is appropriate to study the wide variety of these properties which will
allow to formulate the experimental criteria for the search of this anomalous
``odd''---superconductivity.\ Some work in this direction has already been
done\cite{LJ,KS,KSE,DLS},\ though it was mainly in the direction of comparison
with the properties of high-temperature superconducting oxides.
At the same time there is still no discussion of a number of traditional
problems of the theory of superconductivity as applied to this odd-gap case.

The aim of the present work is to present the microscopic derivation of the
Ginzburg-Landau expansion for the odd-paring state and analyze the behavior
of a number of main characteristic properties of a superconductor near the
transition temperature $T_{c}$.\ We also give the complete analysis of the
upper critical field $H_{c2}$ temperature dependence (we shall see that the
odd superconductors practically always demonstrate the strong type-II
behavior).\ We shall discover a number of anomalies which may be useful for
the experimental search of systems with odd-pairing.\ At the same time the
results obtained give some additional arguments against the use of this model
for the explanation of high-temperature superconductivity in
copper oxides.

The model under discussion is based on the fact\cite{MA},\ that the
weak-coupling BCS gap equation ($N(0)$ - is the density of states at the
Fermi level):
\begin{equation}
\Delta(\xi)=-N(0)\int\limits_{-\infty}^{\infty}d{\xi'}V(\xi,\xi')
\frac{\Delta(\xi')}{2\sqrt{\xi'^{2}+\Delta^{2}(\xi')}}
th\left(\frac{\sqrt{\xi'^{2}+ \Delta^{2}(\xi')}}{2T}\right) \label{1}
\end{equation}
possess the non-trivial solution of the form
$\Delta(\xi)=-\Delta(-\xi)$ (i.e.\ odd over $k-k_{F}$,\ $\xi=v_{F}(k-k_{F})$)
in case of strong enough attractive part of $V(\xi,\xi')$ even if the strong
(or even infinite) point-like repulsive part also exists.\ It is easy to see
\cite{MA},\ that for the odd $\Delta(\xi)$ the repulsive part of interaction in
(\ref{1}) just drops out,\ and the attractive part leads to a pairing state
with non-trivial properties:\
the gap function $\Delta(\xi)$ is zero at the Fermi level leading to the
gapless superconductivity.\ It should be stressed that this pairing state is
isotropic and the energy gap is actually zero everywhere at the Fermi surface,\
which is different from the case of anisotropic pairing e.g of the $d$-wave
type.

In the following we always assume that the interaction kernel in Eq.(\ref{1})
consists of two parts ($E_{F}$ - Fermi energy):
$V(\xi ,\xi ')=V_{1}(\xi ,\xi ')+V_{2}(\xi ,\xi ')$,\ where
\begin{equation}
V_{1}(\xi ,\xi ')=\left\{
\begin{array}{lcl}
U>0&\mbox{ for }&|\xi|,|\xi '|<E_{F} \\
0&\mbox{ for }&|\xi|,|\xi '|>E_{F}
\end{array}
\right.
\label{V1}
\end{equation}
---is the point-like repulsion,\ while $V_{2}(\xi ,\xi ')$---is an effective
pairing interaction (attraction),\ which is nonzero for
$|\xi|,|\xi '|<\omega_{c}$ É $|\xi -\xi '|<\omega_{c}$ (the last property
is of principal importance),\ where $\omega_{c}\ll E_{F}$---is the
characteristic frequency of Bosons,\ responsible for pairing interaction.
Pairing ``potential'' $V_{2}(\xi ,\xi ')$ can be represented by different
model forms\cite{MA}. In the present paper we consider the following
model interaction\cite{KS,KSE}:
\begin{equation} V_{2}(\xi ,\xi ')=\left\{
\begin{array}{lcl}
-Vcos\left(\frac{\pi}{2}\frac{\xi -\xi '}{\omega_{c}}
\right)&\mbox{ for}&|\xi|,|\xi '|,|\xi -\xi '|<\omega_{c}
\\ 0&\mbox{ for }&|\xi|,|\xi '|\mbox{ or}|\xi -\xi '|>\omega_{c}
\end{array}
\right.
\label{V2-cos}
\end{equation}
In this case the integral gap equation reduces to a transcendental one and is
easily solved\cite{KS,KSE},\ while all the main properties characteristic of
the other types of model interactions survive\cite{MA}.

Transition temperature can be obtained from linearized equation:
\begin{equation}
\Delta(\xi)=-N(0)\int\limits_{-\infty}^{\infty}d{\xi '} V(\xi ,\xi ')
\frac{\Delta(\xi ')}{2\xi '} th\left(\frac{\xi '}{2T_{c}}\right)
\label{5}
\end{equation}
It is easily seen\cite{KSE},\ that $T_{c}$ for the odd-gap state is now
determined by:
\begin{equation}
1=g\int\limits_{0}^{\omega_{c}}d{\xi}sin^{2}\left(\frac{\pi}{2}
\frac{\xi}{\omega_{c}}\right)\frac{1}{\xi}th\left(\frac{\xi }{2T_{c}}\right)
\label{16}
\end{equation}
where $g=N(0)V$ --- is the dimensionless coupling constant of the pairing
interaction.\ In this model there exists a critical coupling constant value
and the odd-pairing state appears only for
$g>g_{c}\approx 1.213$,\ even if the repulsive interaction $U$ is absent.
In fact,\ for the weak repulsion the usual ``even'' pairing always dominate
and the appropriate transition temperature is always higher than $T_{c}$ for
the odd pairing.\ As repulsion is increased and for large enough $g$
the odd-pairing becomes energetically favorable\cite{KSE}.
In the following we assume that our system belongs to the odd-paring region on
the phase diagram on the $g$,\ $\mu=N(0)U$ --- plane.

The gap function in our model has the following form\cite{KSE}:
\begin{equation} \Delta(\xi)=\left\{
\begin{array}{lcl}
\Delta_{0}(T)sin\left(\frac{\pi}{2}\frac{\xi}{\omega_{c}}\right)&\mbox{ for
}&|\xi|<\omega_{c}\\ 0&\mbox{ for }&|\xi|>\omega_{c}
\end{array} \right.
\label{17}
\end{equation}
and the temperature dependence of $\Delta_{0}(T)$ is determined by:
\begin{equation}
1=g\int\limits_{0}^{\omega_{c}}d{\xi}sin^{2}\left(\frac{\pi}{2}
\frac{\xi}{\omega_{c}}\right)\frac{th\biggl(\frac{\sqrt{\xi^{2}+
\Delta_{0}^{2}(T)sin^{2}\left(\frac{\pi}{2}\frac{\xi}{\omega_{c}}\right)}}
{2T}\biggr)}{\sqrt{\xi^{2}+\Delta_{0}^{2}(T)sin^{2}\left(\frac{\pi}{2}
\frac{\xi}{\omega_{c}}\right)}}
\label{18}
\end{equation}
The temperature dependence of $\Delta_{0}(T)$ is similar to that of the BCS
theory but does not coincide with it\cite{KSE}.

Normal (non-magnetic) impurities strongly suppress the odd
pairing\cite{KS,KSE}.
In this case the transition temperature is determined by:
\begin{equation}
1=g\int\limits_{0}^{\omega_{c}}\frac{d\xi}{\xi}sin^{2}\left(\frac{\pi}{2}
\frac{\xi}{\omega_{c}}\right)\int\limits_{-\infty}^{\infty}\frac{d\omega}{\pi}
th\left(\frac{\omega+\xi}{2T_{c}}\right)\frac{\gamma}{\omega^{2}+\gamma^{2}}
\label{27}
\end{equation}
where $\gamma$ is the impurity scattering rate. Superconductivity is
completely destroyed for $\gamma\sim T_{c0}$,\ where $T_{c0}$ --- is transition
temperature in the absence of impurities determined by Eq.(\ref{5}).
The critical scattering rate $\gamma_{c}$ leading to the destruction of
superconducting state is determined by the following equation\cite{KS,KSE}:
\begin{equation}
1=\frac{2}{\pi}g\int\limits_{0}^{\omega_{c}}\frac{d\xi}{\xi}sin^{2}\left(
\frac{\pi}{2}\frac{\xi}{\omega_{c}}\right)arctg\left(\frac{\xi}
{\gamma_{c}}\right)
\end{equation}
For $g\simeq g_{c}$ it follows that $\gamma_{c}\sim (g-g_{c})\rightarrow
0$,\ which leads to the corresponding narrowing of superconductivity region
at the phase diagram on the $T_{c}$,$\gamma$---plane.\ In this sense normal
impurities suppress odd-gap superconductivity even faster than magnetic
impurities in the usual ``even'' case.

Ginzburg-Landau theory allows to analyze wider range of physical properties
than the BCS approach due to its limitation to the temperature region close
to transition temperature:\ $T\sim T_{c}$.\ Thus it is important to have the
microscopic derivation of the coefficients of Ginzburg-Landau expansion which
will immediately lead us to a number of important conclusions.

\newpage

\section{Ginzburg-Landau Expansion}

\subsection{Superconductor without impurity scattering}

The gap function (\ref{17}) can be used as an order parameter
in Ginzburg-Landau expansion.\ We assume that its amplitude $\Delta_{0}(T)$
is in general a slowly varying function of spatial coordinates.\ Accordingly
in momentum space we get the Fourier-component of the order parameter:
\begin{equation}
\Delta(\xi,T,{\bf q})=\Delta_{{\bf q}}(T)
sin\left(\frac{\pi}{2}\frac{\xi}{\omega_{c}}\right)
\label{OP}
\end{equation}
and the Ginzburg-Landau expansion for the free-energy-density difference
between
superconducting and normal state in the region of small $q$ takes the form:
\begin{equation}
F_{s}-F_{n}=A|\Delta_{{\bf q}}|^{2}+q^{2}C|\Delta_{{\bf q}}|^{2}
+\frac{1}{2}B|\Delta_{{\bf q}}|^{4}
\label{GL}
\end{equation}
Our task is to find microscopic expressions for the coefficients
$A$,\ $B$ É $C$.

Ginzburg-Landau expansion can be easily derived from diagrams shown in Fig.1.
This is a usual loop-expansion for the free-energy of electrons moving in the
field of superconducting order parameter fluctuations of the form given in
Eq.(\ref{OP}).\ Some additional comments are only needed concerning the
second diagram in Fig.1.\ Its form just guarantees the zero of coefficient
$A$ in GL-expansion (\ref{GL}) at the transition point $T=T_{c}$.
All calculations are standard,\ we only have to take into account the closeness
of temperature $T$ to transition temperature $T_{c}$.\ Some details can be
found in Appendix A.\ Finally we can express GL-coefficients in the following
form:
\begin{equation}
A=A_{0}K_{A},\   B=B_{0}K_{B},\    C=C_{0}K_{C} \label{ABC}
\end{equation}
where $A_{0}$,\ $B_{0}$ and $C_{0}$ are just the usual GL-coefficients for
``even'' pairing\cite{Genn}:
\begin{eqnarray}
A_{0}=N(0)\frac{T-T_{c}}{T_{c}} \\
B_{0}=N(0)\frac{7\zeta(3)}{8\pi^{2}T_{c}^{2}} \\
C_{0}=N(0)\frac{7\zeta(3)}{48\pi^{2}}\frac{v_{F}^{2}}{T_{c}^{2}}
\approx N(0)\xi_{0}^{2}
\label{AGKoeff}
\end{eqnarray}
while everything specific to the odd-pairing is contained in dimensionless
combinations $K_{A}$,\ $K_{B}$ É $K_{C}$:
\begin{equation}
K_{A}=\int\limits_{0}^{\omega_{c}/T_{c}}dx
\frac{sin^{2}(\frac{\pi}{2}\frac{T_{c}}{\omega_{c}}x)}
{2ch^{2}(x/2)}
\label{Ka}
\end{equation}
\begin{equation}
K_{B}=\frac{4\pi^{2}}{7\zeta(3)}\int\limits_{0}^{\omega_{c}/T_{c}}dx
\frac{sin^{4}(\frac{\pi}{2}\frac{T_{c}}{\omega_{c}}x)}{x^2}\left[\frac{th(x/2)}
{x}-\frac{1}{2ch^{2}(x/2)}\right]
\label{Kb}
\end{equation}
\begin{equation}
K_{C}=\frac{1}{2}-\frac{4}{7\zeta(3)}\sum_{n=0}^{\infty}\frac{exp[-\pi^2
\frac{T_{c}}{\omega_{c}}(2n+1)]}
{(2n+1)^3}\left[1+\pi^2\frac{T_{c}}{\omega_{c}}(2n+1)+
\frac{\pi^4}{2}\frac{T_{c}^2}{\omega_{c}^2}(2n+1)^2\right]
\label{Kc}
\end{equation}
These dimensionless expressions are functions of the ratio
$T_{c}/\omega_{c}$,\ which in its turn is the known\cite{KS,KSE} function of
the pairing coupling constant $g$.\ Numerical data for the dependencies of
$K_{A}$,\ $K_{B}$ É $K_{C}$ on $g$ are shown in Fig.2.\ Thus,\ while in
traditional case GL-coefficients depend on $g$ only through the appropriate
dependence of $T_{c}$,\ in our case there appears a new nonmonotonous
dependence of these coefficients on the pairing coupling constant.

Ginzburg-Landau equations define,\ as usual,\ two characteristic lengths:\
the coherence length and penetration depth\cite{Genn}.\ The coherence length
at a given temperature $\xi(T)$ represents the characteristic scale of
inhomogeneities of the order parameter $\Delta$,\ i.e. in fact it is
determined by the ``size'' of the Cooper pair:
\begin{equation}
\xi^2(T)=-\frac{C}{A}
\label{Coh}
\end{equation}
In traditional superconductors:
\begin{eqnarray}
\xi_{BCS}^2(T)=-\frac{C_{0}}{A_{0}} \\
\xi_{BCS}(T)\approx 0.74\frac{\xi_{0}}{\sqrt{1-T/T_{c}}}
\label{CohBCS}
\end{eqnarray}
where $\xi_{0}=0.18v_{F}/T_{c}$.\ In our case we get:
\begin{equation}
\frac{\xi^2(T)}{\xi_{BCS}^2(T)}=\frac{K_{C}}{K_{A}}
\label{Cohodd}
\end{equation}
The dependence of this ratio upon $g$ is shown in Fig.3.

For penetration depth in traditional superconductor we have:
\begin{equation}
\lambda_{BCS}(T)=\frac{1}{\sqrt{2}}\frac{\lambda_{0}}{\sqrt{1-T/T_{c}}}
\label{lambda}
\end{equation}
where $\lambda_{0}^2=\frac{mc^2}{4\pi ne^2}$ determines the penetration depth
for $T=0$.\ In general case we have the following expression for penetration
depth via GL-coefficients\cite{Genn}:
\begin{equation}
\lambda^2(T)=-\frac{c^2}{32\pi e^2}\frac{B}{AC}
\label{GLambda}
\end{equation}
Then for the model under discussion:
\begin{equation}
\frac{\lambda(T)}{\lambda_{BCS}(T)}=\left(\frac{K_{B}}{K_{A}K_{C}}\right)^{1/2}
\label{lambdodd}
\end{equation}
Appropriate dependence upon the coupling constant $g$ is shown in Fig.4.
The divergence of $\lambda$ for $g\rightarrow g_{c}$ is again due to the
disappearance of the odd pairing.

Consider now Ginzburg-Landau parameter:
\begin{equation}
\kappa=\frac{\lambda(T)}{\xi(T)}=\frac{c}{4eC}\sqrt{B/2\pi}
\label{GLpar}
\end{equation}
It is well known that depending on the value of $\kappa$ superconductors are
divided into two classes:\ the values of $\kappa<1/\sqrt{2}$ correspond
to type-I superconductors,\ while $\kappa>1/\sqrt{2}$ define type-II.
In our model of the odd pairing:
\begin{equation}
\frac{\kappa}{\kappa_{BCS}}=\frac{\sqrt{K_{B}}}{K_{C}}
\label{Kappa}
\end{equation}
where
\begin{equation}
\kappa_{BCS}=\frac{3c}{\sqrt{7\zeta(3)}e}\frac{T_{c}}{v_{F}^2\sqrt{N(0)}}
\label{KapBCS}
\end{equation}
---is the Ginzburg-Landau parameter for traditional case.\ The dependence of
$\kappa/\kappa_{BCS}$ on the coupling constant $g$ is shown in Fig.5.\
 From Fig.5 it is clear that for all sensible values of $g$ close to $g_{c}$
the odd-gap superconductor is definitely of type-II.\ Note that the region
of large $g\gg g_{c}$ is actually nonphysical because all our estimates are
based upon weak coupling BCS-like equations like Eq.(\ref{1}),\ while the
correct analysis of the crossover into the strong coupling region requires
\cite{KSE} more serious discussion in the spirit of Ref.\cite{NS}.
Such an analysis in case of the odd-gap pairing has not been performed to
our knowledge.

\subsection{Non-magnetic impurities}

Consider a superconductor containing ``normal'' (nonmagnetic) impurities.
During the derivation of Ginzburg-Landau expansion we have to take into
account now impurity scattering processes shown in diagrams like Fig.6(a,b).
It is easy to convince ourselves that the contribution of diagrams shown
in Fig.6(b) is actually zero which is connected with vertex contributions
being odd over the variable $\xi$ (factors of $sin(\xi)$).\ Thus, the loop
expansion for the odd-gap superconductor with impurities acquires the form
shown in Fig.6(c,d) (up to the second-order terms),\ where electronic lines
denote average Green's functions renormalized by impurity scattering.\ There
is no ``diffusion'' renormalization due to diagrams like shown in Fig.6(b),\
(impurity ``ladder'') which is characteristic of the usual theory of ``dirty''
superconductors\cite{G}.\  In this respect the structure of all expressions
in our model is closer to that of usual theory of the ``clean'' limit.\
Note,\ however,\ that the usual ``dirty'' limit $\xi_{0}\ll l$ (where $l$ ---
is the mean-free path),\ cannot actually be reached for the odd-gap
superconductors because this pairing state is completely destroyed by
impurity scattering already for $\gamma\sim T_{c}$\cite{KS,KSE}.\ Below we
shall discuss the major changes of GL-coefficients $A$ É $C$ due to impurity
scattering.\ Details again are to be found in Appendix A.

Ginzburg-Landau coefficients are again represented by Eqs.(\ref{ABC}),\
while impurity scattering renormalizes the dimensionless functions $K_{A}$ and
$K_{C}$,\ which now take the following form:
\begin{equation}
K_{A}^d=\frac{T_{c0}}{T_{c}}\int\limits_{0}^{\omega_{c}/T_{c0}}\frac{dx}{x}
sin^{2}(\frac{\pi}{2}\frac{T_{c0}}{\omega_{c}}x)
\int\limits_{-\infty}^{\infty}\frac{dy}{2\pi}\frac{y+x}{ch^{2}\left(\frac{y+x}
{2}\frac{T_{c0}}{T_{c}}\right)}\frac{\gamma/T_{c0}}{y^2+\gamma^2/T_{c0}^2}
\label{Kad}
\end{equation}
\begin{eqnarray}
K_{C}^d=\frac{4\pi^3}{7\zeta(3)}\frac{T_{c}}{T_{c0}}\sum_{n=0}^{\infty}
\Biggl\{\frac{1}{[(2n+1)\pi \frac{T_{c}}{T_{c0}}+\frac{\gamma}{T_{c0}}]^3}-
\frac{exp(-\pi\frac{T_{c0}}{\omega_{c}}[(2n+1)\pi\frac{T_{c}}{T_{c0}}+
\frac{\gamma}{T_{c0}}])}{[(2n+1)\pi \frac{T_{c}}{T_{c0}}+\frac{\gamma}
{T_{c0}}]^3}\times \\
\times\left[1+\pi\frac{T_{c0}}{\omega_{c}}[(2n+1)\pi \frac{T_{c}}{T_{c0}}+
\frac{\gamma}{T_{c0}}]+\frac{\pi^2}{2}\frac{T_{c0}^2}{\omega_{c}^2}
[(2n+1)\pi\frac{T_{c}}{T_{c0}}+\frac{\gamma}{T_{c0}}]^2 \right]\Biggr\}
\label{Kcd}
\nonumber
\end{eqnarray}
Here $T_{c0}$ --- is transition temperature in the absence of impurity
scattering,\ while $T_{c}$ --- is real transition temperature in the impure
system,\ which is determined by Eq.(\ref{27}),\ and $\gamma$ --- is impurity
scattering rate.\ In the limit of $\gamma\rightarrow 0$ Eqs.(\ref{Kad}) and
(30) naturally reduce to Eqs.(\ref{Ka}) and (\ref{Kc}).

Numerically we obtain the dependencies of $K_{A}^d/K_{A}$ and $K_{C}^d/K_{C}$
on the impurity scattering rate $\gamma$,\ shown in Fig.7 and Fig.8.\
Most important is the dependence of $K_{A}^d$,\ which rapidly drops to zero as
$\gamma\rightarrow \gamma_{c}$.

GL-coefficients $A$ and $C$,\ as usual,\ define the temperature dependence of
the upper critical field close to $T_{c}$:
\cite{Genn}:
\begin{equation}
H_{c2}(T)=\frac{\phi_{0}}{2\pi\xi^2(T)}=-\frac{\phi_{0}}{2\pi}\frac{A}{C}
\label{Hc}
\end{equation}
where $\phi_{0}=c\pi/e$ --- is magnetic flux quantum.\ Now we can easily find
the ``slope'' of the temperature dependence of $H_{c2}(T)$ near $T_{c}$,\ i.e.
the temperature derivative:
\begin{equation}
|\frac{dH_{c2}}{dT}|_{T_{c}}=\frac{24\pi\phi_{0}}{7\zeta(3)v_{F}^2}T_{c}\frac
{K_{A}^d}{K_{C}^d}
\label{Hcslope}
\end{equation}
In Fig.9 we show the dependence of normalized derivative $|dH_{c2}/dT|_{T_{c}}$
on disorder (scattering rate $\gamma$).\ It is seen that the slope of $H_{c2}$
rapidly diminishes and drops to zero as disorder grows and
$\gamma\rightarrow \gamma_{c}$.\ This behavior is just opposite to that of the
usual theory,\ where in the ``clean'' limit the slope of $H_{c2}$ does not
depend on impurity concentration,\ while in the ``dirty'' limit it grows with
impurity scattering\cite{G}.\ This anomalous behavior may be used as an
experimental criterion for the search of superconductors with odd-gap
pairing state.\ Note that for high-temperature superconducting oxides such a
behavior is not observed and the odd-gap model is apparently unable to
explain the anomalies of $H_{c2}$ observed in these systems under
disordering\cite{Nova}.\ This adds to previous arguments against the use of
the odd-gap pairing model to describe superconductivity in copper oxides
\cite{KS,KSE}.

\newpage

\section{Upper critical field}

In view of the anomalies of $H_{c2}$ behavior in the odd-gap superconductor
close to $T_{c}$ it is of some interest to perform a complete analysis of
temperature dependence of the upper critical field in these systems taking
into account the effects of impurity scattering.\ This analysis is to be done
on full microscopic basis,\ i.e.\ via the study of Cooper instability in an
external magnetic field.\ In the absence of magnetic field the Cooper
instability leading to odd-gap pairing was analyzed in Ref.\cite{KSE}.\
Below we use a similar approach and include impurity scattering from the very
beginning.

Diagrams describing both impurity scattering and pairing interaction in Cooper
channel are shown in Fig.10.\ Fig.10(a) represents impurity scattering vertex
in ``ladder'' approximation,\ while at Fig.10(b) we introduce the vertex part
which includes pairing interaction (represented by dot).\ It can be obtained
from the following integral equation:
\begin{equation}
\Phi_{{\bf pp'}}({\bf q}\omega\omega')=\Gamma_{{\bf pp'}}({\bf q}\omega)
\delta_{\omega\omega'}-T\sum_{\omega_{1}}\sum_{\bf p_{1}p_{2}}
\Gamma_{{\bf pp_{1}}}({\bf q}\omega)V({\bf p_{1}p_{2}})\Phi_{\bf p_{2}p'}({\bf
q}\omega_{1}\omega')
\label{Phi}
\end{equation}
where the impurity scattering vertex
$\Gamma_{{\bf pp'}}({\bf q}\omega)$ is defined by the ``ladder'' equation of
Fig.10(a) and $V({\bf pp'})$ --- is our pairing interaction.\
Following Ref.\cite{KSE} we introduce the following vertex parts,\ summed
over Matsubara frequencies:
\begin{equation}
\Gamma_{{\bf pp'}}({\bf q})=-T\sum_{\omega}\Gamma_{{\bf pp'}}({\bf q}\omega)
\label{Gm}
\end{equation}
\begin{equation}
\Phi_{{\bf pp'}}({\bf q})=-T\sum_{\omega\omega'}\Phi_{{\bf pp'}}({\bf q}
\omega\omega')
\label{Ph}
\end{equation}
which satisfy the equation following from (\ref{Phi}):
\begin{equation}
\Phi_{{\bf pp'}}({\bf q})=\Gamma_{{\bf pp'}}({\bf q})
-T\sum_{\bf p_{1}p_{2}}
\Gamma_{{\bf pp_{1}}}({\bf q})V({\bf p_{1}p_{2}})\Phi_{\bf p_{2}p'}({\bf q})
\label{Phs}
\end{equation}
Cooper instability leading to the odd-gap pairing can be found from equation
shown in Fig.10(c) which in fact represents the response function for
fluctuations of our (odd-gap) order parameter:
\begin{equation}
\Pi({\bf q})=\Pi_{0}({\bf q})-\Pi_{0}({\bf q})V\Pi({\bf q})
\label{Pie}
\end{equation}
where the loop graphs represent
\begin{equation}
\Pi_{0}({\bf q})=\sum_{|\xi_{{\bf p}}|,|\xi_{{\bf p'}}|<\omega_{c}}
sin\left(\frac{\pi}{2}\frac{\xi_{{\bf p}}}{\omega_{c}}\right)\Gamma_{{\bf pp'}}
({\bf q})sin\left(\frac{\pi}{2}\frac{\xi_{{\bf p'}}}{\omega_{c}}\right)
\label{Pi0}
\end{equation}
\begin{equation}
\Pi({\bf q})=\sum_{|\xi_{{\bf p}}|,|\xi_{{\bf p'}}|<\omega_{c}}
sin\left(\frac{\pi}{2}\frac{\xi_{{\bf p}}}{\omega_{c}}\right)\Phi_{{\bf pp'}}
({\bf q})sin\left(\frac{\pi}{2}\frac{\xi_{{\bf p'}}}{\omega_{c}}\right)
\label{Pi}
\end{equation}
and we have taken into account the explicit form of pairing interaction in our
model(\ref{V2-cos}).
Solution of Eq.(\ref{Pie}) is obviously:
\begin{equation}
\Pi({\bf q})=\frac{\Pi_{0}({\bf q})}{1+V\Pi_{0}({\bf q})}
\label{Pii}
\end{equation}
Divergence of this expression (zero of the denominator --- divergence of the
response function) defines the Cooper instability.

It is easy to see that due to the odd nature of vertexes over the variable
$\xi$ only the first graph of the ladder in Fig.10(a) contribute to
$\Pi_{0}({\bf q})$ while the diffusion contribution vanishes.\ Then we have:
\begin{eqnarray}
\Pi_{0}({\bf q})=-T\sum_{\omega}\sum_{|\xi|<\omega_{c}}sin^2\left(\frac
{\pi}{2}\frac{\xi}{\omega_{c}}\right)G_{\omega}({\bf p+q}/2)G_{-\omega}
({\bf p-q}/2)= \\ \nonumber
=-TN(0)\sum_{\omega}\int\limits_{-\omega_{c}}^{\omega_{c}} d\xi
sin^2\left(\frac{\pi}{2}\frac{\xi}{\omega_{c}}\right)
\frac{1}{2}\int\limits_{-1}^{1}dt\frac{1}
{\xi-v_{F}qt/2+i(\omega+\gamma)}\frac{1}{\xi+v_{F}qt/2-i(\omega+\gamma)}= \\
\nonumber
=-TN(0)\sum_{\omega}\int\limits_{-\omega_{c}}^{\omega_{c}} d\xi
sin^2\left(\frac{\pi}{2}\frac{\xi}{\omega_{c}}\right)
\frac{1}{4\theta\xi}
ln\left(\frac{(\theta+\xi)^2+\Omega^2}{(\theta+\xi)^2-\Omega^2}\right)
\nonumber
\end{eqnarray}
where $\Omega=\omega+\gamma$ and $\theta=v_{F}q/2$.\ Thus the divergence of
(\ref{Pii}) is determined by the equation:
\begin{equation}
1=gT\sum_{\omega}\int\limits_{-\omega_{c}}^{\omega_{c}} d\xi
sin^2\left(\frac{\pi}{2}\frac{\xi}{\omega_{c}}\right)\frac{1}{4\theta\xi}
ln\left(\frac{(\theta+\xi)^2+\Omega^2}{(\theta+\xi)^2-\Omega^2}\right)
\label{Tthet}
\end{equation}
which for $q=0$ reduces to:
\begin{equation}
1=gT\sum_{\omega}\int\limits_{-\omega_{c}}^{\omega_{c}}d\xi
\frac{sin^2(\frac{\pi}{2}\frac{\xi}{\omega_{c}})}{\xi^2+(\omega+\gamma)^2}
\label{Tcimp}
\end{equation}
 From here by standard manipulations (cf.\ (\ref{sum})) we get Eq.(\ref{27}).

For a system in an external magnetic field $H$ the Cooper pair momentum
${\bf q}$ is replaced as usual by ${\bf q}-\frac{2e}{c}{\bf A}$,\ where ${\bf
A}$--- is vector-potential.\ In this case the Cooper instability is again
determined by Eq.(\ref{Tthet}),\ only $\theta$,\ now denotes
$\theta=v_{F}q_{0}/2 $,\ where $q_{0}$---is the minimal eigenvalue
of an operator $({\bf q}-\frac{2e}{c}{\bf A})^2$,\ which is obviously equal
\cite{Genn} to $\sqrt{2\pi\frac{H}{\phi_{0}}}$,\ where $\phi_{0}$ -
is the magnetic flux quantum introduced above and corresponding to double
electronic charge.\ In this way we easily obtain the equation determining
the upper critical field at arbitrary temperature.

Numerical solution of this equation is conveniently done after reducing the
discrete sum over Matsubara frequencies to integral.\ Details can be found in
Appendix B.\ As a result,\ instead of Eq.(\ref{Tthet}) we obtain the following
equation for $H_{c2}$:
\begin{equation}
1=2gT\int\limits_{-\omega_{c}}^{\omega_{c}}d\xi sin^2\left(\frac{\pi}{2}
\frac{\xi}{\omega_{c}}\right)\int\limits_{0}^{\infty} dx  sin\left(\frac
{v_{F}x\sqrt{\frac{\pi H}{2\phi_{0}}}}{2\pi
T}\right)\frac{exp[-(\frac{\gamma}{2\pi T}+\frac{1}{2})x]
sin(\frac{\xi x}{2\pi T})}{v_{F}\sqrt{\frac{\pi H}{2\phi_{0}}}\xi x(1-exp(-x))
}
\label{Hc2T}
\end{equation}
In Fig.11 we present the results of numerical solution of Eq.(\ref{Hc2T}) for
different scattering rates $\gamma$.\ Qualitative form of temperature
dependence of $H_{c2}$ is more or less usual,\ but we can clearly see rather
rapid drop of the slope of $H_{c2}(T)$-curve close to $T_{c}$ as scattering
increases.\ This drop is in completely described by expressions obtained above
within Ginzburg-Landau approach.\ Note that the qualitative form of $H_{c2}$
is practically the same for different values of the coupling constant $g$,\
thus on Fig.11 we show the data for only one value of this constant.\
We should like to stress again that this anomalous behavior of $H_{c2}$ with
the growth of disorder is characteristic of the odd-gap pairing and can be
used in experimental search of this ``exotic'' state.

\newpage
\section{Conclusion}

The main part of this paper was devoted to microscopic derivation of the
coefficients of Ginzburg-Landau expansion in the model of pairing state with
energy gap odd over $k-k_{F}$.\ We have shown that an additional strong
dependence of GL-coefficients on the pairing coupling constant appears,\ which
does not reduce to the usual coupling constant dependence of $T_{c}$.\
We have analyzed this coupling constant dependence for the main parameters of
Ginzburg-Landau theory.

The main theoretical conclusion following from our analysis of impurity
scattering in the odd-gap model reduces to the absence of ``diffusion''
renormalization,\ which leads to the dimensional dependencies of
GL-coefficients
characteristic of ``clean'' superconductors.\ At the same time a rapid drop of
$T_{c}$ due to impurity scattering and important dependence of GL-coefficients
on impurity scattering rate lead to anomalous behavior of $H_{c2}$ with
disordering which is quite different both from usual ``clean'' limit behavior
or from traditional ``dirty'' case.\ This anomalous behavior is reflected in
rather rapid drop of the slope of $H_{c2}(T)$-curve close to $T_{c}$ with
increased impurity scattering which can be used as an experimental criterion in
the search of systems with odd-gap pairing.

We have also analyzed the temperature behavior of $H_{c2}$ for all relevant
temperatures,\ solving for the odd-gap Cooper instability in an external
magnetic field.\ The problem is again characterized by the absence of
``diffusion'' renormalization,\ which is due again to the odd over $k-k_{F}$
nature of the order parameter.\ Results obtained are in complete correspondence
with Ginzburg-Landau approach.

This work was partially supported by the Scientific Council on High-
Temperature Superconductivity under the project $N^{o}$ 93-001 and also by
Soros Foundation grant RGL000, as well as by the grant the Russian
Foundation of Fundamental Research $N^{o}$ 93-02-2066 .

\newpage

\begin{center}
{\bf APPENDIX A}
\end{center}

Next follow some details of calculations leading to our expressions for
GL-coefficients.\ Diagram shown in Fig.1(a) gives the following expression
with terms up to second order in $q$:
\begin{eqnarray}
Fig.1(a)=-|\Delta_{\bf q}|^2\frac{T}{(2\pi)^3}\sum_{\omega}\int {d{\bf p}}
sin^2\left(\frac{\pi\xi}{2\omega_{c}}\right)G_{\omega}({\bf p+q}/2)
G_{-\omega}({\bf -p+q}/2)\approx \\ \nonumber
\approx -|\Delta_{\bf q}|^{2}N(0)\Biggl\{\int\limits_{0}^{\omega_{c}}d\xi
sin^2\left(\frac{\pi\xi}{2\omega_{c}}\right)\frac{1}{\xi}th\frac{\xi}{2T}+ \\
\nonumber
+\frac{\pi v_{F}^{2}q^{2}}{48}T \sum_{\omega}\frac{1}{|\omega|^3}\left[2-
2exp\left(-\pi\frac{|\omega|}{\omega_{c}}\right)\left(1+\pi\frac{|\omega|}{\omega_{c}}+
\pi^2\frac{|\omega|^2}{2\omega_{c}^2}\right)\right]\Biggr\}
\label{fig1a}
\end{eqnarray}
Here $G_{\omega}({\bf p})=[i\omega - \xi_{\bf p}]^{-1}$ is the usual Matsubara
Green's function of an electron,\ $\omega=(2n+1)\pi T$.\
In the integrand we have also to expand up to first order in powers of
$T-T_{c}$
and in terms containing the small parameter $q^2$,\ we just put $T=T_{c}$.\
Using the equation for $T_{c}$ (\ref{16}) it is easily seen that the
contribution of diagram of Fig.1(b) is:
\begin{equation}
Fig.1(b)=-|\Delta_{\bf q}|^2\Biggl\{\frac{T}{(2\pi)^3}\sum_{\omega}\int d{\bf
p}
G_{\omega}({\bf p})G_{-\omega}(-{\bf p})sin^2\left(\frac{\pi\xi}{2\omega_{c}}
\right)\Biggr\}=-\frac{N(0)}{g}|\Delta_{\bf q}|^2
\label{fig1b}
\end{equation}
which compensates the zero-order terms over $T-T_{c}$ and $q$ in the
expression for Fig.1(a).\  As a result we get the expressions given in
(\ref{ABC}),\ (\ref{Ka}) É (\ref{Kc}),\ which define coefficients $A$ and $C$.

For the impure superconductor all calculations are similar,\ we only have to
take into account that Green's function now is equal to: $G_{\omega}({\bf p})=
[i\omega - \xi_{\bf p} + i\gamma sign(\omega)]^{-1}$.\ Accordingly the
expression for diagram in Fig.6(d) contains the sum:
\begin{equation}
T_{c}\sum_{\omega}\frac{1}{(\omega+\gamma)^2+\xi^2}=\int\limits_{-\infty}
^{\infty}\frac{d\omega}{\pi}\frac{1}{2\xi}th\left(\frac{\omega+\xi}{2T_{c}}\right)
\frac{\gamma}{\omega^2+\gamma^2}
\label{sum}
\end{equation}
and reduces,\ after the use of $T_{c}$-equation (\ref{27}),\ to:
\begin{equation}
Fig.6(d)=-N(0)|\Delta_{\bf q}|^2\int\limits_{0}^{\omega_{c}}\frac{d\xi}{\xi}
sin^2\left(\frac{\pi\xi}{2\omega_{c}}\right)\int\limits_{-\infty}^{\infty}
\frac{d\omega}{\pi}th\left(\frac{\omega+\xi}{2T_{c}}\right)\frac{\gamma}
{\omega^2+\gamma^2}=-\frac{N(0)}{g}|\Delta_{\bf q}|^2
\label{fig6g}
\end{equation}
This contribution cancels terms of zeroth order in  $T-T_{c}$ and $q^2$
in the expression for diagram in Fig.6(c).\ As a result GL-coefficients depend
on real transition temperature $T_{c}$ in the presence of impurities.
Analogously we must take into account impurity scattering in part of
diagram in Fig.6(c) which determines contributions of the order of $T-T_{c}$
and $q^2$.\ Finally we obtain expressions for GL-coefficients $A$ and $C$,\
quoted above for the impure case.

The simplest way to find GL-coefficient $B$ is to use the expression for the
free-energy difference between superconducting and normal phases at arbitrary
temperature for homogeneous order-parameter:
\begin{equation}
F_{s}-F_{n}=N(0)\int\limits_{-\infty}^{\infty} d\xi\Biggl\{\frac{|\Delta(\xi)|
^2}{2\epsilon}th\left(\frac{\epsilon}{2T}\right)-2Tln\left(\frac{ch(\epsilon/2T)}
{ch(\xi/2T)}\right)\Biggr\}
\label{FsFn}
\end{equation}
where $\epsilon=\sqrt{\xi^2+|\Delta(\xi)|^2}$,\ and $\Delta(\xi)$
is defined by Eqs.(\ref{17}) and (\ref{18}).\ Expanding in powers of
$\Delta_{0}$ and $T-T_{c}$,\ and using $T_{c}$-equation (\ref{16}),\ we get:
\begin{eqnarray}
F_{s}-F_{n}=N(0)\frac{\Delta_{0}^2}{g}+N(0)\int\limits_{0}^{\omega_{c}} d\xi
\Biggl\{-\frac{\Delta_{0}^2}{\xi}sin^2\left(\frac{\pi\xi}{2\omega_{c}}\right)
th\left(\frac{\xi}{2T_{c}}\right)+ \\ \nonumber
+(T-T_{c})\Delta_{0}^2 \frac{sin^2\left(\frac{\pi\xi}{2\omega_{c}}\right)}
{2T_{c}^2ch^2\left(\frac{\xi}{2T_{c}}\right)}+\frac{1}{4}\frac{\Delta_{0}^4}{\xi^2} sin^4
\left(\frac{\pi\xi}{2\omega_{c}}\right)\left(\frac{1}{\xi}th\left(\frac{\xi}
{2T_{c}}\right)
-\frac{1}{2T_{c}}\frac{1}{ch^2\left(\frac{\xi}{2T_{c}}\right)}\right)\Biggr\}
\nonumber
\end{eqnarray}
Using $T_{c}$-equation (\ref{16}) again we can isolate the coefficient $A$
of $\Delta_{0}^2$ which coincides with that found above.\ For $\Delta_{0}^4/2$
we obtain the coefficient $B$ as:
\begin{equation}
B=\frac{N(0)}{2}\int\limits_{0}^{\omega_{c}}\frac{d\xi}{\xi^3}sin^4\left(
\frac{\pi\xi}{2\omega_{c}}\right)th\left(\frac{\xi}{2T_{c}}\right)-
\frac{N(0)}{4T_{c}}\int\limits_{0}^{\omega_{c}}\frac{d\xi}{\xi^2}\frac
{sin^4\left(\frac{\pi\xi}{2\omega_{c}}\right)}{ch^2\left(\frac{\xi}{2T_{c}}\right)}
\label{BB}
\end{equation}
which immediately leads to Eq.(\ref{Kb}).

\newpage
\begin{center}
{\bf APPENDIX B}
\end{center}

To calculate $\Pi_{0}({\bf q})$ we must find the following sum:
\begin{equation}
S=\sum_{\omega}\frac{1}{\xi-\theta t+i(\gamma+\omega)}\frac{1}
{\xi+\theta t-i(\gamma+\omega)}
\label{S}
\end{equation}
Convenient representation of this sum via integral can be obtained with the
use of the well-known formula:
\begin{equation}
\sum_{n=0}^{\infty}\frac{1}{n+a}\frac{1}{n+b}=\frac{1}{b-a}[\psi(b)-\psi(a)]
\label{dilog}
\end{equation}
and integral form of logarithmic derivative of the $\Gamma$-function:
\begin{equation}
\psi(z)=\int\limits_{0}^{\infty}dx \left(\frac{e^{-x}}{x}-\frac{e^{-zx}}{1-
e^{-x}}\right) \qquad Re(z)>0
\label{psi}
\end{equation}
so that
\begin{equation}
\psi(b)-\psi(a)=\int\limits_{0}^{\infty}\frac{e^{-ax}-e^{-bx}}{1-e^{-x}}
\qquad Re(a)>0 \quad Re(b)>0
\label{psipsi}
\end{equation}
Then we obtain:
\begin{equation}
S=\frac{1}{\pi T\xi}\int\limits_{0}^{\infty} dx sin\left(\frac{\xi x}
{2\pi T}\right)\frac{exp(-(\frac{\gamma}{2\pi T}+\frac{1}{2})x)
exp(-i\frac{\theta tx}{2\pi T})}
{1-exp(-x)}
\label{SSS}
\end{equation}
Integration over $t$ is elementary and as a result Eq.(\ref{Tthet})
reduces to Eq.(\ref{Hc2T}).

\newpage
\begin{center}
{\bf Figure captions.}
\end{center}

\vskip 0.4cm
Fig.1.\ Diagrammatic form of Ginzburg-Landau expansion. Wavy lines
represent fluctuations of the order-parameter (\ref{OP}).\ Diagram (b) is
calculated for $T=T_{c}$.

Fig.2.\ $K_{A}$,\ $K_{B}$ É $K_{C}$ dependence on coupling constant $g$.

Fig.3.\ Dimensionless coherence length dependence on $g$.

Fig.4.\ Dimensionless penetration depth as a function of $g$.

Fig.5.\ Ginzburg-Landau parameter as a function of coupling constant $g$.

Fig.6.\ Diagrams for free-energy in the presence of impurity scattering.\
Dashed lines --- impurity scattering.

Fig.7.\ Normalized coefficient $K_{A}^d$ as a function of impurity
scattering rate for different values of $g$:

$g:$ 1---1.22;\ 2---1.24;\ 3---1.3;\ 4---1.5;\ 5---2.0;\ 6---5.0;\ 7---10.0

Fig.8.\ Normalized coefficient $K_{C}^d$ as a function of impurity scattering
rate for different values of $g$:

$g:$ 1---1.22;\ 2---1.24;\ 3---1.3;\ 4---1.5;\ 5---2.0;\ 6---5.0;\ 7---10.0

Fig.9.\ The slope of $H_{c2}$-curve close to $T_{c}$ as a function of impurity
scattering rate for different values of $g$:

$g:$ 1---1.22;\ 2---1.24;\ 3---1.3;\ 4---1.5;\ 5---2.0;\ 6---5.0;\ 7---10.0

Derivative of the upper critical field is normalized upon its value in the
absence of impurity scattering.

Fig.10.\ Diagrams for Cooper instability in the impure system:

(a)---vertex part of impurity scattering in Cooper channel.
(b)---vertex part of pairing interaction.
(c)---response function determining odd-gap pairing instability.

Fig.11.\ Temperature behavior of the upper critical field for systems with
different degree of disorder $\gamma/T_{c0}$.\
Pairing coupling constant $g=2$,\
magnetic field in units of
$H_{0}=\frac{2}{\pi}\frac{\phi_{0}T_{c0}}{v_{F}^2}$,\
temperature normalized by $T_{c}$,\ depending on disorder.

$\gamma/T_{c0}$: 1---0;\ 2---0.25;\ 3---0.5;\ 4----0.75;\ 5---0.87

\end{document}